\documentstyle{article}
\input{psfig.tex}
\title{Defect models of cosmic structure in light of the new CMB
data\footnote{To appear in the proceedings of the XXXVth Rencontres de 
Moriond ``Energy Densities in the Universe''.}
}
\author{Andreas Albrecht\\ U. C. Davis Department of Physics\\
           One Shields Avenue, Davis, CA, 95616}
\begin{document}

\maketitle

\centerline{\bf Abstract}
Defect models of cosmic structure formation have faced a number of difficult 
challenges over recent years. Yet interestingly, new CMB data does not show 
strong evidence for secondary oscillations in the anisotropy power. 
Here I review the current standing of the cosmic defect models of 
structure formation in light of the current data.

\section{Introduction}

The idea that a network of defects seeded the formation of cosmic
structure has received intensive study, and for a long period was
viewed as the only viable competitor for the popular inflation-based
models.  This idea has intrinsic appeal due to its basis in 
spontaneous symmetry breaking (one of the central features of modern
particle physics) and because the correct {\em amplitude} of the cosmic
structure can be naturally produced. (The right amplitude requires a
Grand Unification scale for spontaneous symmetry breaking, a very
natural prospect in high energy physics.)  

More recently the
defect models have faced a number of 
serious challenges in the face of data that increasingly favor the
inflation-based models.  However, recent results from the
Boomerang98\cite{Boom98} 
and MAXIMA\cite{MAXIMA} experiments do not show strong secondary
peaks in the 
Cosmic Microwave Background (CMB) anisotropy power.  The lack of these
secondary oscillations has been established as a strong signature of
cosmic defect models.  It thus is natural to ask whether it is time to
revisit the cosmic defect models.  In this article I review the
difficulties that face the defect models and discuss possible
work-arounds. (The reader might also be interested in the
recent reviews by Durrer\cite{Durrer:00} and Magueijo and
Brandenberger\cite{MagueijoBrandenberger:00} which provide a complementary 
discussion, and the excellent book by Vilenkin and Shellard\cite{VS} for
background information.)  
I conclude that Boomerang98 and MAXIMA data actually
add to the case {\em against} the defect models, and that radical
variations from the standard picture (or even from the already radical 
attempts to save the defect models) would be required if a
defect-based model were to eventually give the favored description
of the formation of cosmic structure.

\section{Past challenges}

\subsection{Overview}
Cosmic defect models are notoriously difficult to
calculate because, unlike inflation-based models, they in principle
require one to track highly non-linear behavior from very early epochs
(essentially from the Grand Unification epoch until today).  However,
early researchers in this area were encouraged by the pioneering work
of Kibble\cite{Kibble1} in which he argued that networks of cosmic defects might
undergo a simple scaling behavior which would greatly simplify the
required calculations.

The original numerical simulations focused on local cosmic
strings\cite{AlbrechtTurok1,AlbrechtTurok2}. The first local cosmic string simulations supported the idea
of scaling, but then subsequent higher resolution simulations called
the earlier results into question\cite{Allen1,BennettBouchet1}. The
new simulations 
always had very important dynamics occurring right at the scale of
resolution, and these dynamics affected the behavior of the entire
string network.  This result violated the simple scaling picture,
which held that one scale (proportional to the Hubble radius)
dominated the defect dynamics at all times.  While there seemed to be
a physical basis for these multi-scale dynamics (namely the buildup of 
kinks on the string), the inability of the simulation results to
decouple from processes at the limits of the numerical resolution cast 
doubt on the applicability of the numerical simulations to cosmology.

By contrast, ``global'' defects are the result of the breaking of
global symmetries, and the defects are thus necessarily coupled to
a massless particle (the Goldstone Boson).  This coupling appears to
damp out the small-scale dynamics sufficiently for global defects that 
their scaling behavior has been observed and has not
been called into question. 

Without scaling arguments it is truly impossible to achieve the
dynamic range necessary for a calculation to confront the whole array
of modern
cosmological data.  Thus, the global defects are at a significant
practical advantage.  However, arguments have also been made that the
coupling of local strings to gravity, while weaker, ultimately has a
similar effect to that of the Goldstone Boson, and creates a feedback
mechanism that imposes scaling behavior after a sufficiently long
period of time\cite{AustinEtal:93}.  A new generation of local defect
simulations have been constructed that (like the originals) exhibit
scaling behavior\cite{ACDKSS}.  This 
is not because they have incorporated gravitational back-reaction, but 
because they (again like the originals) have a sufficiently unprecise
evolution of small scale features that the scaling behavior is
recovered.  The justification for this has been the idea that a full
treatment including gravitational back-reaction would result in
scaling behavior. A very interesting alternative approach used by
Copeland {\it et al.}\cite{CopelandEtal:00} has been to
use extensions\cite{AustinEtal:93} to Kibble's original
phenomenological treatment\cite{Kibble1}  

The history of attempts to calculate cosmic structure formation from
defect models is a long one.  While the earliest work looks remarkably
simplistic by current standards, important tools were developed
that are still in use today.  I start my discussion with the more
modern work that came out in 1997, when a number of groups produced
complementary results using somewhat different approaches.  

Allen {\it et al.}\cite{ACDKSS} performed calculations based entirely
on numerical 
simulations.  Their 
strength was that the  fewest additional assumptions needed to be
made, but their weakest point was that with limited dynamic range,
many important questions could not be addressed.  My collaboration\cite{ABR}
was at the other extreme.  Over the years my concern had been growing
that there were so many uncertainties surrounding numerical
simulations that it was important to take a more flexible approach,
where the impact of the uncertainties could be fed through to the
final result.  The  
possible uncertainties ranged from numerical uncertainties in the
simulations to the fact that a great variety of defect models were
possible.  There was a growing tendency (among non-experts) to assume
that a single defect simulation could be representative of the whole
range of possible defect models.  Our collaboration used parameterized
forms for the defect two-point functions and imposed scaling ``by
hand'' (a method originated by Albrecht and
Stebbins\cite{AS}) to achieve 
dynamic range.  Our strength was that at last some of the defect
uncertainties could be explored in a systematic way, but our weakness
was that it was not clear that any given choice of our parameters
would reproduce a given defect model exactly.  In fact, we relied
considerably on results from Allen {\it et al.}\cite{ACDKSS} to set crucial
parameters in our 
models. A similar approach was taken by Durrer and
Kunz\cite{DurrerKunz:97} who focused on modeling global
defects. Turok's collaboration lay in between these two
extremes\cite{PenEtal:97}.  All  
the defect two-point functions were calculated from simulations, and
then scaling arguments were used to produce dynamic range.  Because
they studied {\em global} defects, the scaling assumption was well
justified.  All the groups found their results showed serious problems 
when confronted with the data.  These problems are the subject of the
next subsection:

\subsection{EDS Universe with ``perfect'' scaling}
\begin{figure}
\centerline{\psfig{figure=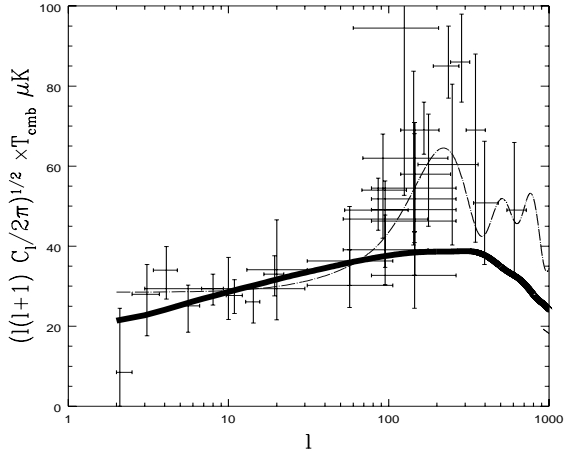,width=3in,height=2.5in}}
\caption{The heavy curve shows a typical defect prediction for the CMB 
anisotropy power in an EDS model with perfect scaling. This figure is
from Albrecht {\it et al.}\protect\cite{ABR} and also shows data of that
time and the standard Cold 
Dark Matter model}
\label{EDSCMB}
\end{figure}

The major publications of 1997 all considered defects in an
Einstein De Sitter (EDS) universe
and ``perfect'' scaling.
The most obvious problem that was observed by all the groups was the 
serious lack of power at moderate scales in the CMB anisotropies, when 
normalized to COBE. Figure \ref{EDSCMB} illustrates the problem.
Even with the relatively poor data of that time the defect models
appeared to be in trouble.
At this stage, many people outside the field came away with the
impression that defect models were in trouble because of the low CMB
power.  

However, Albrecht {\it et al.}\cite{ABR} and Durrer {\it et
al.}\cite{DurrerEtal:97} examined the effect of the various
uncertainties on this conclusion.  Our results showed that 
very modest deviations from perfect scaling 
would result in significant power in the CMB at the relevant
scales. We emphasized the point 
that many defect networks are known to experience a transient deviation from
scaling during the radiation-matter transition (a point that has been
known for a long time\cite{Kibble1}).  We found that modest
uncertainties about this 
transient were enough to prevent the CMB power from being a problem
for defect models at that time.  
(I note here that such transients are
much less likely to occur in global defect models, because again the
back-reaction from Goldstone Bosons drives the network more firmly
toward the scaling solution.)  Durrer {\it et
al.}\cite{DurrerEtal:97} explored other uncertainties and also
concluded that the CMB power could be increased.  As an explicit illustration
of the potential effect of of numerical uncertainties, contrast the
results of Contaldi {\it et al.}\cite{ContaldiEtal:98} with those of
Allen {\it et al.}\cite{ACDKSS}

The real problem for the EDS models was the matter power spectrum.
When the perturbations were COBE normalized, the matter power was
completely wrong.  This result was so dramatic that there had been
hints of this problem in earlier work, but the 1997
calculations were the first to have sufficient dynamic range to
confront this problem head-on.  The problem became known as the ``b100
problem'' because it was on scales of 100 Mpc where the matter power
spectrum performed particularly badly (the predicted power was many
times lower than the observations, as illustrated in
Fig. \ref{matter}).  Earlier work might have revealed this problem
more clearly if it had not focused on the matter power on 8 Mpc (a
traditional point of comparison) which was not nearly as problematic as the
power on 100 Mpc. The problem was so extreme, that few expected
possible non-Gaussian effects could provide a resolution.

\subsection{$\Lambda$ models with Radiation-Matter transients}
\label{Lmwrmt}
\begin{figure}
\centerline{\psfig{figure=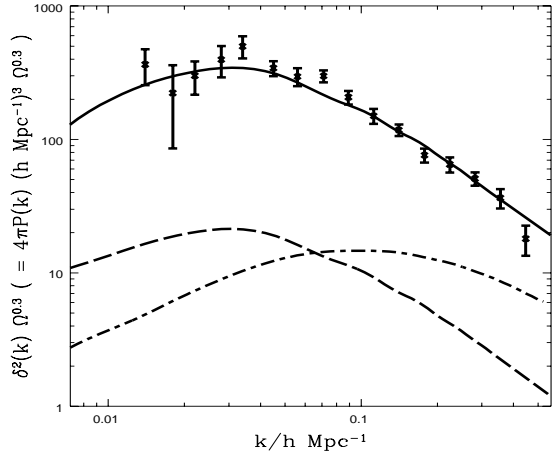,width=3in,height=2.5in}}
\caption{Matter power from the $\Lambda$ defect model (solid curve)
using a realistic radiation-matter transient, and 
assuming a bias factor of 2. The dot-dashed curves show the matter
power for a perfectly scaling EDS string model. The dashed curve is
for the $\Lambda$ universe with perfect scaling. (These last two have
the bias set to unity.)
The curves are from Battye {\it et al.}\protect\cite{BAR:97} and are all
COBE normalized.} 
\label{matter}
\end{figure}

The uncertainties in the defect models that we explored were
unable to significantly reduce the b100 problem\cite{ABR}.  However, we soon
realized that things were quite different for models with a cosmological
constant \cite{BAR:97}.  For these models the scales affected by the
defects during the radiation-matter transition affected different
scales today, and it turned out for $\Omega_{\Lambda}=0.7$ the
transition was perfectly placed for transients to resolve the b100
problem, as illustrated in Fig. \ref{matter}.  
In fact, we used a fit to the transients actually observed
in one of the simulations \cite{MSa} to produce a model that
looked very realistic.  The building evidence that there is a cosmological
constant (or something similar) added support for this model, and for
a while cosmic defects were again in the running as a realistic
picture of cosmic structure formation.

\section{The current picture}

\begin{figure}
\centerline{\psfig{figure=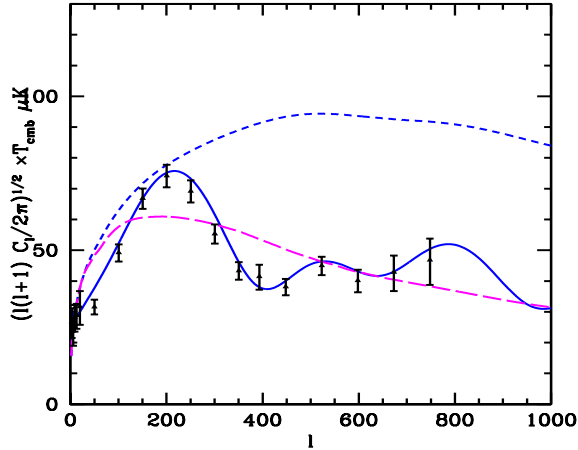,width=3in,height=2.5in}}
\caption{Current data (as complied by Knox\protect\cite{radpack}) with
two defect models 
(dashed) and an inflation-based model (solid). The upper defect model
has a standard ionization history and the lower model has an
ionization history specifically designed to produce a sharper, shifted peak.}
\label{now}
\end{figure}

Figure \ref{now} shows a compilation of the current data
\cite{radpack} along with 
two (dashed) curves from defect-based models, and one (solid) curve
from a ``best fit'' inflation-based model\cite{radpack}.  
The upper defect curve is from the ``successful'' model discussed in
Section \ref{Lmwrmt}. Clearly in the face of new CMB data this model is 
in trouble.  While the upper defect curve assumed the standard
ionization history, the lower curves use a non-standard ionization
history specifically designed to shift the peak in the curve
\cite{WellerEtal:99}. 

Considering that the lower defect curve was calculated well before the
new data came in, it is striking how it appears to give a best
``smooth fit'' to the data.  However, there is no question that the
inflation-based curve gives a much better fit to the same data.  Using 
RADPACK \cite{radpack} and using the COBE, Boomerang98 and MAXIMA data, 
(46 data points) we get $\chi^2 = 132.5$ for the best defect model,
and $\chi ^2 = 47.9$ for the inflation-based model.  
The new data has re-focused the problem back onto the CMB, were
defect models have great difficulties reproducing the observed spectrum.

\section{How much trouble are defects models in?}

\subsection{Coherence}
\label{C}
Probably the most fundamental distinguishing difference between
inflation-based models and defect models of cosmic structure is
connected with ``coherence''.  Passive models of structure formation
have a primordial spectrum of perturbations imposed at some very early 
time, which then experiences an extended period of {\em linear}
evolution.  Inflation-based models of structure formation are examples 
of passive models \cite{ACFM,AAPrince,AANATO}.  Active models, such as the
defect models, have
non-linear processes figuring significantly in structure formation 
throughout time.  The oscillating behavior (``Sakharov'', or ``acoustic''
oscillations) exhibited by the curve from the inflationary model in
Fig. \ref{now} is a characteristic of passive models, and no realistic 
active model has exhibited any oscillatory structure in the CMB
anisotropy power\footnote{
There is one active model, the ``mimic
model''\protect\cite{mimic} which achieves a peak structure just like that of
a passive model (although the signal is {\em not} identical in the
CMB polarization power).  However, the mimic model requires the
non-linear matter to exhibit extraordinary coherence itself, and to press 
hard against the limits set by causality.  No one has offered a clear
picture of how the mimic model could be realized.
}.

Considerable attention has been given to the suggestion in the new
data that the 2nd acoustic peak is absent, or suppressed, compared
with inflation models that are favored for other reasons\cite{peak2}.
It is natural to ask whether this is an indication that defect models
should be coming back into favor.  Unfortunately for the defect
models, the coherence issues are already showing up strongly in the
first peak, which clearly {\em does} exist in the  data.  In particular, no
realistic defect model has been able generate a first peak with
anywhere near the sharpness exhibited in the
data\footnote{Contaldi\protect\cite{Contaldi:00} has speculated that
defects could produce a sharp peak in a universe with supercritical
density.  However, this approach fails by producing large deficiencies
in anisotropy power for $l > 350 $. }. 

The fundamental problem is that realistic defect models do not affect
the cosmic structure on one scale with a single easily isolated defect 
motion.  If this were possible, perhaps the defect physics could be
specially designed to created peaks and dips in the CMB anisotropy
power (it was necessary to assume this could be done to construct the
``mimic model'' referred to in footnote {\it a}) . The known defect
models make contributions to a given scale of cosmic structure from
many different defect motions occurring at many different times.
Furthermore, the non-linearity of the defects gives the defect motion
an effectively ``random'' component. This
tends to wash out any attempts one might make to create a sharp peak.
Also the work of Weller {\it et al.}\cite{WellerEtal:99} attempted to
doctor the ionization history (quite artificially) to produce a sharp
first peak, but the degree of sharpness exhibited by real acoustic
oscillations could not be duplicated. The result of our best efforts
can been seen in Fig. \ref{now}.  

So within the familiar scope of defect behavior, the shape of the
observed CMB anisotropy power appears to be impossible to achieve.
The contrast with the success of the passive (coherent)  models in
generating the 
first peak is certainly striking, and thus  we are already
seeing a serious failure of the defect models due to their decoherent
behavior. Any concerns that might exist within the inflationary
paradigm regarding the apparently low 2nd peak at this stage appear
much more minor.  

\subsection{Scaling}
Recall that in most calculations the scaling property (or modest
deviations from it) was put in by hand. In some cases there was only
indirect evidence that scaling was a realistic assumption.  Could
certain cosmic defect models save themselves by dramatically violating the
scaling assumption?  The problem here is that overall a scaling
spectrum of perturbations is just what is needed to account for the
observed cosmic structure over many scales. It appears that any
violation of scaling would have to be localized specifically to
generate a peak in the CMB anisotropy power, rather than be a general
property of the defect evolution.  Then we are back facing the problem 
mentioned in Section \ref{C}, that the impact of the defects is not
highly time-localized, so it is unlikely  that a sharp peak
could be produced by any ``glitch'' in the defect network.

\subsection{Numerical Uncertainties}
There are many places where numerical uncertainties
significantly detract from our understanding of cosmic defects,
especially in the case of local defects where the numerical
work has hardly converged.  However, much has been done to model the
possible numerical uncertainties, and the problems defect models have
in fitting the apparent first acoustic peak do not seem to be
resolved by assuming any particular type of numerical error.

\subsection{Filling out the picture}
The reader may have noticed that the above discussion uses words like
``appears'' and ``seems'' extensively.  Exactly why am I not taking a
more concrete stand?  The fact is that despite the progress and
simplification that has been made over the last several years, working 
out the predictions from defect models remains a pretty 
complicated business.  Instead of saying ``it is {\em unlikely} that that
a sharp peak could be produced by any `glitch' in the defect network'' 
it would be nice to be able to say that all possible glitches have
been checked and it is 100\% certain that none could produce a sharp
peak.  Similar comments apply to other discussions throughout this
section.

The fact is that given the effort involved, and the declining fortunes 
of the defect models in the face of new data, there is less and less
interest in undertaking the effort to check out all possibilities.  Of
course that situation could change if the passive models run into
serious problems.

It should also be clear from this paper that my years of experience
have built up certain prejudices about how defects behave.  Perhaps
someone free of these prejudices will discover some new defect
behavior out there in the vast non-linear world of the different types 
of defects.  This remains to be seen.

One last comment on this topic:  When modeling different possible
defect behaviors, one of the great problems is coming up with models of 
the defect stress-energy that are consistent with known constraints
(especially those imposed by causality).  If someone were starting
from scratch to explore the possibilities, my advice would be to use
the excellent techniques developed by Pen {\it et al.}\cite{PenEtal:97}. Pen {\it
et al.} applied their techniques to calculate the implications of
particular numerical simulations, but the same techniques could also
be combined with the parameterized approach similar to those of
Albrecht {\it et al.}\cite{ABR}. I suspect this would be the most
effective way of moving forward.

\subsection{Other roles for defects} 
\label{orfd}
\begin{figure}
\centerline{\psfig{figure=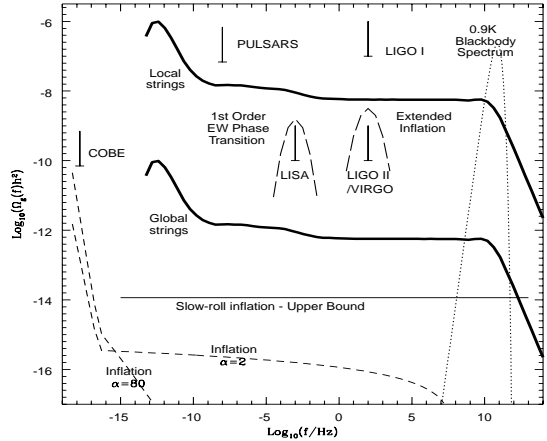,width=3in,height=2.5in}}
\caption{This summary figure from a review by Battye and
Shellard\protect\cite{BattyeShellard} shows the possible gravity wave signals
from two different cosmic defect models. The failure of defect models
of structure formation puts a small downward pressure on the defect
curves, but still allows interesting possibilities for a concrete
detection. } 
\label{gwaves}
\end{figure}

The entire discussion thus far has been about models where the
universe started in a completely homogeneous state and defects were
solely responsible for the formation of cosmic structure.  I have
concluded that all known models of this type have failed to reproduce
the observed cosmic structure.  However, reducing the mass scale of
the defects slightly would make them completely irrelevant for cosmic
structure, but they could still have many interesting observable effects as
sources of cosmic rays, gravitational waves\cite{BattyeShellard} (see Fig. 
\ref{gwaves}), or as players 
in the out of equilibrium processes that produce baryon
asymmetry\cite{Davis:99,MagueijoBrandenberger:00}. The 
only things that have been ruled out are the defect models of cosmic structure
formation.  

There also has been much discussion recently of the ``middle ground'', 
where the defects have a partial role in cosmic structure\cite{hybrid}.  While my
gut reaction is quite negative to this idea, the reason for this
reaction is purely due to the prejudice that nature should do things
in a simpler way.  If nature goes through all the trouble to produce
passive perturbations, why bother adding defects to the mix?  I
suppose this argument can be countered by the fact that if defects are 
formed by symmetry breaking at the grand unification scale (a likely
enough prospect), they will naturally have masses {\em around} values
that will contribute noticeably to structure formation without
requiring that they account for everything.  Fortunately, on this 
 point we should eventually be in a position where experiment,
rather than prejudice, determines the outcome.

\section{Conclusions}
In this brief review a have addressed three different aspects of the
defect models of cosmic structure.  First of all, I have shown how the 
defect models, as they have long been understood to behave, are clearly
ruled out by the modern CMB data.  Then I
discussed whether some gap in our understanding
could mean that the defect models are not really in as much trouble as 
they appear to be.  I have argued that it is hard to see how any of
the uncertainties in defect scenarios could come around to save
defect-based structure formation.  I have also tried to be frank about 
the limitations inherent to this type of argument.
Finally, I have emphasized that there are a host of other potentially
interesting cosmological effects from defects. The failure of
defect models of structure formation only serves to place a modest
constraint on the overall amplitude of these other effects.

I wish to thank Yong-Seon Song for calculating the $\chi^2$'s using
RADPACK and Ben Gold for comments on this manuscript.  This work was
supported by DOE grant DE-FG03-91ER40674, and UC
Davis.

\end{document}